\documentclass[useAMS,usenatbib]{mn2e}

\usepackage{graphicx}
\usepackage{amssymb}
\usepackage{slashbox}
\usepackage{threeparttable}
\newcommand{\eqb}{\begin{eqnarray}}
\newcommand{\eqe}{\end{eqnarray}}
\newcommand{\sth}{\sigma_{\rm T}}

\newcommand{\dt}{\delta t}

\newcommand{\Lgtot}{L_{\gamma}^{\rm tot}}
\newcommand{\lp}{\ell_{\rm p}^{\rm inj}}

\newcommand{\rb}{r_{\rm b}}
\newcommand{\eb}{\epsilon_{\rm B}}

\newcommand{\gmx}{\gamma_{\max}}

\newcommand{\gp}{\gamma_{\rm p}}
\newcommand{\mpr}{m_{\rm p}}
\newcommand{\mel}{m_{\rm e}}
\newcommand{\hp}{\eta_{\rm p}}
\newcommand{\ep}{\epsilon_{\rm p}}
\newcommand{\nph}{n_{\gamma}}

\title[GRB neutrinos and hadronic cascades]
{The role of hadronic cascades in GRB models  of efficient neutrino production}
\author[M. Petropoulou ]{Maria Petropoulou$^{1,2}$ \thanks{E-mail:
mpetropo@purdue.edu} 
 \\
$^1$Department of Physics and Astronomy, Purdue University, 525 Northwestern Avenue, West Lafayette, IN, 47907, USA \\
$^2$ NASA Einstein Postdoctoral Fellow}
\begin{document}
\date{Received.../Accepted...}

\pagerange{\pageref{firstpage}--\pageref{lastpage}} \pubyear{2014}

\maketitle

\label{firstpage}
\begin{abstract}
We investigate the effects of hadronic cascades on the gamma-ray burst (GRB) prompt emission spectra
in scenarios of efficient neutrino production. By assuming 
a fiducial GRB spectrum and a power-law proton distribution extending to ultra-high energies,
we calculate the proton cooling rate and the neutrino emission produced through photopion processes. For this, we employ
a numerical code that follows the formation of the hadronic cascade by taking into account 
non-linear feedback effects, such as the evolution of the target photon field itself due
to the contribution of secondary particles. We show that in cases of efficient proton cooling and subsequently efficient high-energy neutrino production, the emission
from the hadronic cascade distorts and may even dominate the GRB spectrum. Taking this into account,
we constrain the allowable values of the ratio $\eta_{\rm p}=L_{\rm p}/L_{\gamma}$, where $L_{\rm p}$ and $L_{\gamma}$ are the isotropic equivalent proton and prompt
gamma-ray luminosities. For the highest value of $\eta_{\rm p}$ that does not lead to the dominance of the cascading emission,
 we then calculate the maximum neutrino luminosity from a single burst and show that it ranges between  $(0.01- 0.6)L_{\rm p}$ 
 and $(0.5-1.4)L_{\gamma}$ for various parameter sets.
 We discuss possible implications of other parameters, such as the magnetic field strength and the shape of the initial gamma-ray spectrum, on our results.
 Finally, we compare the upper limit on $\hp$ derived here with various studies in the field, and we 
 point out the necessity of a self-consistent treatment of the hadronic emission in order to avoid erroneously high
 neutrino fluxes from GRB models.
\end{abstract}

\begin{keywords}
neutrinos -- radiation mechanisms: non-thermal -- gamma ray burst: general 
\end{keywords}
\section{Introduction}
Gamma-ray bursts (GRBs) are candidate sites of proton acceleration to ultra-high energies (UHE) \citep{waxman95, vietri95} and, therefore,
potential sources of high-energy (HE) neutrino emission \citep{paczynskixu94, waxmanbahcall97}.  
The problem of GRB neutrino production has been 
considered by many authors \citep{murase08, mannheim01, dermeratoyan03, guetta04, asano05, rachenmeszaros98, baerwald13, reynoso14} and 
it may gain even more interest under the light of the recent IceCube HE neutrino detection \citep{icecube13, aartsen14}. 

Another aspect of GRB models for neutrino production is the formation of hadronic cascades (HC), i.e.
cascades consisting of relativistic electron-positron pairs, 
initiated by photopion interactions of UHE protons with the GRB prompt 
radiation which, in this framework, serves as the target field. 
The emission produced by such cascades has  some interesting implications as far as the prompt emission is concerned.
For example, it has been proposed as an alternative explanation
for the underlying power-law components seen in some bright bursts (e.g. GRB 090902B \citep{abdo09}; GRB 080319B \citep{racusin08}), which
extend from the hard X-rays up to GeV energies and do not agree with simple extrapolations of the MeV spectrum \citep{asanoinoue10}.
More general studies regarding the emission signatures of hadronic cascades in the GeV and TeV energy bands and how
these can be used as diagnostic tools of UHECR acceleration in GRBs have been made by various authors (e.g. \citealt{boettcherdermer98, asanoinoue07,asanoetal09}).
One common feature of the aforementioned studies is that the emission produced by the hadronic cascade
does not dominate over the MeV emission of the burst, which further implies that the proton cooling and neutrino production 
through photopion interactions are not very efficient. 
On the other hand, models that focus on the GRB neutrino emission require high efficiency while at the same time they neglect 
any effects of the hadronic cascade on the MeV part of the GRB spectrum (e.g. \citealt{abbasi10, zhangkumar13}). 

In the present work we try to bridge the gap between the two approaches by demonstrating,
in the most general way possible, 
the gradual dominance of the cascading emission over the MeV (and/or GeV) part of the spectrum as
the efficiency in neutrino production progressively increases.
For this, we simulate the GRB prompt emission spectrum as a grey-body photon field, 
which can be characterized only by two parameters, i.e.
its effective temperature $T$ and its compactness $\ell_{\gamma}$. 
Using this as a fixed target field, we gradually increase the injection luminosity 
of protons, or equivalently, their compactness, up to that value
where the emission from the HC
begins to affect
the gamma-ray spectrum. 
The dominance of the HC sets, therefore, an upper limit on $L_{\rm p}$ and on the ratio $\hp=L_{\rm p}/L_{\gamma}$,
where $L_{\rm p}$ and $L_{\gamma}$ are the proton and prompt gamma-ray (isotropic) luminosities, respectively, as measured in the observer's frame.
For the maximum value of this ratio ($\eta_{\rm p,\max}$) we then calculate the neutrino production efficiency ($\xi_{\nu}$) as well as the efficiency in the injection 
of secondary pairs and photons ($\xi_{\rm sec}$) into the HC. 
We show that the upper limit of $\hp$ is anticorrelated with the $\ell_{\gamma}$, $\xi_{\nu}$ and $\xi_{\rm sec}$.
As a next step, we investigate the robustness of our findings by repeating the procedure for a higher magnetic field strength
and a gamma-ray spectrum that can be described by  the more accurate for GRBs Band function \citep{band09}.
We comment also on the role of the Bethe-Heitler process in the formation of the HC.
Finally, we calculate $\ell_{\gamma}$ for different parameter sets used in the literature and
compare the values of `proton loading' used therein, i.e. the ratio of proton to gamma-ray luminosities (or energy densities), 
with our upper limit $\eta_{\rm p, \max}$. 

The present work is structured as follows: in \S 2 we describe our methods and we then present 
the numerical code used for our simulations in \S 3. We continue in \S 4 with the presentation of our results and 
discuss the effects of other parameters, such as the spectral shape of the initial gamma-ray emission.
In \S 5 we compare our upper limit imposed on the `proton loading' with that used in various cases in the literature.
We conclude in \S 6 with a summary and a discussion of our results.

\section{Methods}
In the present study we do not attempt a self-consistent calculation of the GRB prompt emission nor 
we pinpoint the GRB emission itself. Instead, we focus on the formation of the hadronic cascade
and the self-consistent calculation of the neutrino emission after taking into account the 
modification of the initial gamma-ray spectrum.
We choose, first, to depict the prompt emission spectrum as a grey-body photon field instead of using 
the typical Band function for two reasons: the 
 effects of the cascade emission on the primary photon field become more evident and only two parameters, namely
the photon compactness $\ell_{\gamma}$ and the effective temperature $T$, are required for its description. The effects of a Band-shape
photon spectrum on our results will be discussed separately in \S 4.4.2.

We assume that at a distance $r$ from the central engine 
protons are accelerated into a power-law distribution with index $p$ to ultra-high energies (UHE), e.g.
$E_{\rm p} < 10^{18}$~eV in the comoving frame, and are subsequently injected
at a constant rate into a spherical region\footnote{The assumption of a spherical region is valid as long as
the beaming angle $1/\Gamma$ is smaller than the opening angle of the jet, which holds
during the GRB prompt phase.} of size $\rb$ that moves outwards from the central engine
with Lorentz factor $\Gamma$. This region is equivalent to
the shell of shocked ejecta in the internal shock scenario, and 
has a comoving width $\rb \simeq r/\Gamma$ (for reviews, see \citealt{piran04, zhangmeszaros04}).

We further assume that the region contains a magnetic field of strength $B$ (in the comoving frame),
which is usually  related to the jet  kinetic luminosity $L_{\rm j}$ through the parameter $\epsilon_B$ as follows:
\eqb
\eb L_{\rm j} = c B^2 \Gamma^2 r^2.
\label{B}
\eqe
Protons with gyroradii larger than the size $\rb$ cannot be confined, and thus
escape from the region \citep{hillas84}. This sets a maximum Lorentz factor that is given by
\eqb
 \gamma_{\rm H}= \frac{e B \rb}{\mpr c^2}.
\label{gpmx}
\eqe
In principle, the maximum proton energy is given by the minimum of $\gamma_{\rm H}$ and $\gamma_{\rm sat}$, where
the latter denotes the saturation energy of the acceleration process due to energy losses. 
For the parameters used throughout the text, we find
that $\gamma_{\rm H} \lesssim \gamma_{\rm sat}$ (see Appendix~B). For this, we set $\gmx=\gamma_{\rm H}$. 
The total proton injection luminosity $L_{\rm p}$, which is just a fraction of $L_{\rm j}$, 
can be used for defining the  proton injection compactness as:
\eqb
\lp = \frac{\sth L_{\rm p}}{ 4 \pi \rb \Gamma^4 \mpr c^3}.
\eqe
Although electron acceleration at high energies
is also expected to take place, here, in our attempt to minimize the number of free parameters,
we assume that the injection luminosity of primary relativistic 
electrons is much lower than that of protons, making
their contribution to the overall photon emission negligible.

The following parameters were kept fixed in all our numerical simulations, unless stated otherwise:
$\Gamma=225$, $\rb=10^{12}$~cm,  $B=960$~G, $\gmx=10^{8.6}$ and  $T=10^7$~K. 
From this point on, we will refer to this parameter set as the benchmark case. 
For these values,  the typical pulse duration in the internal shock scenario would
be $\dt \approx \rb / c\Gamma = 0.15$~s, while $\eb \simeq 2\times 10^{-3}$ for $L_{\rm j}=3\times10^{52}$~erg/s.\footnote{The choice
of a low $\epsilon_{\rm B}$ value will be justified later on.}
For  a given pair of $\rb$ and  $\Gamma$, different values of $\ell_{\gamma}$ correspond to different $\gamma$-ray (isotropic) luminosities ($L_{\gamma}$), since
the photon compactness is defined as
\eqb
\ell_{\gamma} = \frac{\sth L_{\gamma}}{4\pi \rb \Gamma^4 \mel c^3}
\eqe
In particular, we chose the following set of $\ell_{\gamma}$ values, namely $\{0.07, 0.22, 0.7, 2.2, 7.0\}$,
which corresponds to prompt gamma-ray luminosities in the range $L_{\gamma}=10^{50} - 10^{52}$~erg/s with a logarithmic step of 0.5.
Then, for each value of $\ell_{\gamma}$, we performed a series of numerical simulations where we increased $\lp$ over its previous value
by a fixed logarithmic step $\delta x$.

The method outlined above highlights the main difference between our approach and the one usually adopted 
in the literature (e.g. \citealt{asanoinoue07,asanoetal09}).
Here, we use as free parameters the gamma-ray and proton injection compactnesses instead of 
$\Gamma$, $r$, $\dt$, $L_{\gamma}$ and $L_{\rm p}$, since only the former are the intrinsic 
 quantities of the physical system. 
Note that very different combinations of $\Gamma, r, \dt, L_{\gamma}$
 may lead to the same $\ell_{\gamma}$ and to similar derived properties of the leptohadronic system, such as the neutrino production efficiency.

\section{Numerical code}
In order to study the formation of the hadronic cascade and its effects on the multiwavelength photon spectrum,
we employ the time-dependent numerical code as presented in \cite{dmpr12} -- hereafter DMPR12.
This follows the evolution of protons, neutrons, secondary pairs, photons and neutrinos by solving the coupled  integro-differential equations
that describe the various distributions. The coupling of energy losses and injection introduces a self-consistency in this approach that 
allows the study of the system at various conditions, e.g. in the presence of non-linear electromagnetic (EM) cascades
(see also \citealt{petromast12b} for a relevant discussion). 
Although details can be found in DMPR12, for the sake of completeness, we summarize in Table~\ref{table0} the physical processes that are included in the code.
\begin{table*}
 \caption{Physical processes that act as injection (source) and loss terms in the kinetic equations of each species.}
 \begin{threeparttable}
 \centering
 \begin{tabular}{l c c c c c c c}  
 \hline
  & protons & neutrons & pions & muons & relativistic pairs & photons & neutrinos \\
  \hline
  Injection & \multicolumn{7}{c}{} \\
  \hline
  \hline
  & external &  $p \gamma$   &  $p \gamma$      &    $p \gamma$     &   $p \gamma$   &  neutral pion decay   & 	$p \gamma$ \\
  &  $p \gamma$\tnote{a}  &        &        &         & BH\tnote{b} \ pair production &     proton synchrotron   & 	 $\beta$- decay \\
  & $\beta$-decay &        &        &         & $\gamma \gamma$ pair production   &   electron synchrotron & 			\\
  & 	          &       &         &        &                                    &  muon synchrotron  &\\
  & 		 &        &         &        &                                    &  inverse Compton & \\
  \hline
  Loss & \multicolumn{7}{c}{} \\
    \hline
  \hline
  & $p \gamma$ & $p \gamma$ 	& decay	&  decay & synchrotron & $\gamma \gamma$ pair production & escape \\
  & BH         & $\beta$-decay &  &  synchrotron & inverse Compton & synchrotron self-absorption &  \\
  & synchrotron& escape &               &   &     annihilation &   Compton downscattering &  \\
  & escape     &        &               &         &    escape        &    escape & \\
  \hline
 \end{tabular}
  \begin{tablenotes}
 \item[a] photopion process
 \item[b] Bethe-Heitler process
 \end{tablenotes}
 \end{threeparttable}
 \label{table0}
\end{table*}
%
%

Photohadronic interactions are modelled using the results of Monte Carlo simulations. In particular, for Bethe-Heitler
pair production the Monte Carlo results by \cite{protheroe96} were used  (see also \citealt{mastetal05}). Photopion interactions
were incorporated in the time-dependent code by using the results of the Monte Carlo event generator SOPHIA \citep{muecke00}, which
takes into account channels of multipion production for interactions much above the threshold.

Synchrotron radiation of muons was not included in the version of the code presented in DMPR12 and for 
the exact treatment we refer the reader to \cite{dpm14}. As synchrotron cooling of pions is not yet included in the 
numerical code, we restricted our analysis to cases
where the effects of pion cooling are minimal. 

Pairs that cool down to Lorentz factors $\gamma \sim 1$ contribute to the Thomson depth and they 
are treated as a separate population with 
$k T_e \ll \mel c^2$ \citep{lightmanzd87}. For the pair annihilation and photon downscattering processes we 
followed \cite{coppiblandford90} and \cite{lightmanzd87}, respectively (for more details see \citealt{mastkirk95}).  
For completeness, we note that the limits on $\hp$ derived in \S 4 would have been even more strict
if photon downscattering was not taken into account, or cooling of pairs down to non-relativistic temperatures was not possible.
\section{Results}
\subsection{Proton energy loss rate}
\begin{figure}
   \centering
    \includegraphics[width=0.5\textwidth]{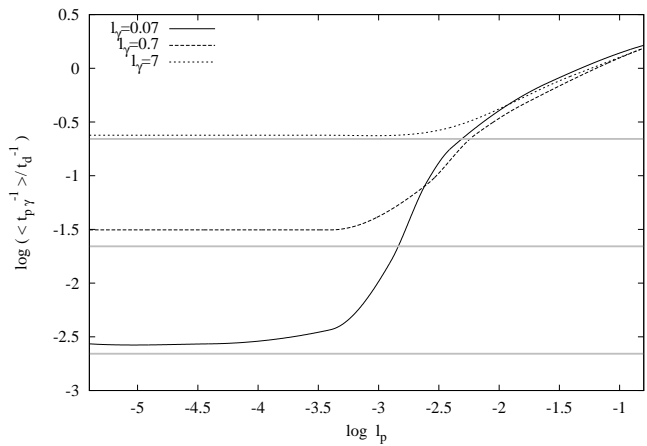}
        \caption{Photopion cooling rate (in units of $c/\rb$) of protons with $\gamma_p =10^6 > E_{\rm th}/1.4 kT \simeq 10^5$, where $E_{\rm th}\simeq 0.15$~GeV, 
        as a function of the proton injection compactness $\lp$ for three values of the gamma-ray compactness marked on the plot. 
        For comparison reasons, the cooling rate given by eq.~(\ref{anal}) is plotted with grey lines. }
        \label{fig1}
   \end{figure}

First, we show that for high enough proton injection compactnesses,
photons produced through the hadronic cascade contribute to the target photon field for photopion interactions and therefore
enhance the respective proton energy loss rate. 
For this, we compare the analytic expression for the energy loss rate 
on a grey-body photon field of certain $\ell_{\gamma}$ and $T$ with the one derived numerically after taking into account
the modification of the photon spectrum because of the cascade. 
It can be shown (for more details, see Appendix~A) that the fractional energy loss rate of 
protons with Lorentz factor $\gp > \gamma_{\rm th} = E_{\rm th}/1.4kT$, where $E_{\rm th}\simeq 0.15$~GeV, is 
\eqb
t_{\rm p\gamma}^{-1} \simeq 5\times10^{-5}\frac{\ell_{\gamma}}{\Theta} t_{\rm d}^{-1},
\label{anal}
\eqe
where $\Theta= kT/\mel c^2$ and $t_{\rm d}=\rb/c$.
As expected (see e.g. WB97, \citealt{aharonian00}),
the above expression does not depend on the proton injection compactness.
The characteristic loss  rate\footnote{The loss rate is calculated by  
$\langle t_{\rm p \gamma}^{-1} \rangle = \int d\gp \gp n_{\rm p}(\gp) \dot{P}_{\rm p\gamma} /  \int d\gp \gp n_{\rm p}(\gp) $, where
$n_{\rm p}$ is the steady-state proton distribution.}, as derived by the simulations,
where the feedback on the target field is taken into account in a self-consisent way, 
is plotted against $\lp$ in Fig.~\ref{fig1} for three values of $\ell_{\gamma}$ marked on the plot. 
In all cases, the energy loss rate is constant
and in agreement with the analytic expression (\ref{anal}), shown with grey lines in Fig.~\ref{fig1}, only up to a certain value of $\lp$. Its subsequent rapid increase
is a sign of the modification of the initial photon spectrum because of the hadronic cascade. Hence, the analytic estimates of
the fraction of energy lost by proton to pions can be considered only as a lower limit.

    \begin{figure}
   \centering
    \includegraphics[width=0.5\textwidth]{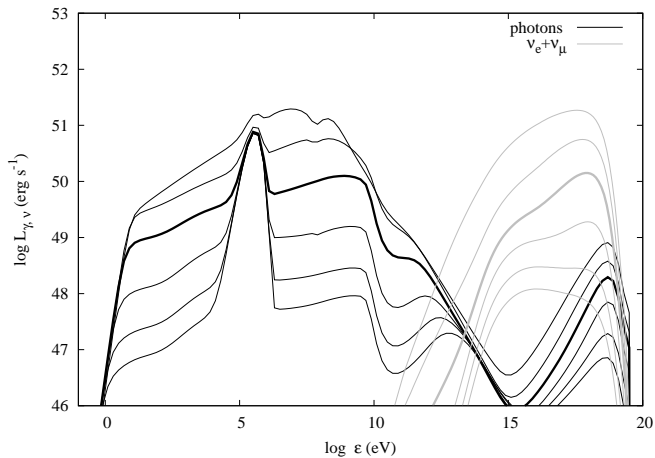}
    \caption{Observed photon and neutrino spectra obtained for $\ell_{\gamma}=0.7$ and values of the proton injection compactness
    starting from $\ell_p=10^{-3.4}$ (bottom curve) up to $\ell_p=10^{-1.4}$ (upper curve) with logarithmic increaments of 0.4. 
    The last spectra
    before the dominance of the hadronic cascade are shown with thick lines. } 
    \label{sed07}
   \end{figure}
    \begin{figure}
   \centering
    \includegraphics[width=0.5\textwidth]{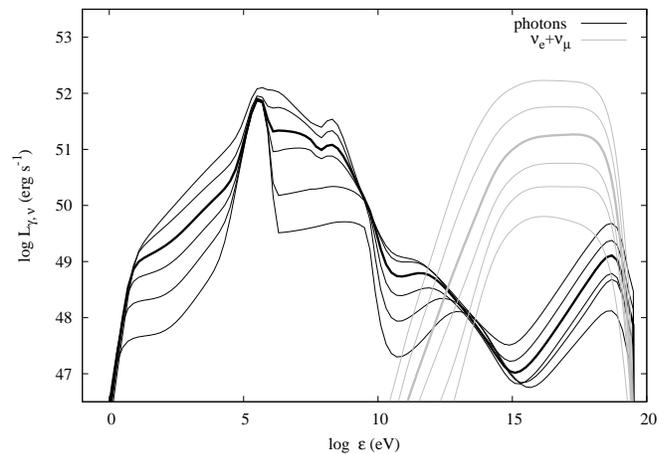}
    \caption{Observed photon and neutrino spectra obtained for $\ell_{\gamma}=7$ and values of the proton injection compactness
    starting from $\ell_p=10^{-2.6}$ (bottom curve) up to $\ell_p=10^{-0.6}$ (upper curve) with logarithmic increaments of 0.4. Thick lines
    have the same meaning as in Fig.~\ref{fig1}.}
    \label{sed7}
   \end{figure}
\subsection{Photon and neutrino emission}
The modification of the photon spectrum due to the emission from the secondaries produced in the hadronic cascade is 
demonstrated in Figs.~\ref{sed07} and \ref{sed7}, where we plot the multiwavelength photon spectra (black lines)
 in the observer's frame
for two indicative values of the gamma-ray compactness, namely 
$\ell_{\gamma}=0.7$ and $7$. For our choice of $\Gamma$ and $\rb$ (see \S 2) these correspond to observed
gamma-ray luminosities $L_{\gamma}=10^{51}$ and $10^{52}$~erg/s, respectively. The neutrino ($\nu_{\rm e}+\nu_{\mu}$) spectra (grey lines)
are also plotted, for comparison reasons.
In both figures, each spectrum is obtained by increasing $\lp$ (or  equivalently $L_{\rm p}$) over its previous value with a logarithmic step $\delta x$ --
for the exact values see labels of the respective figures. 

Above a certain value of $\lp$, which depends on the value of $\ell_{\gamma}$,  
the emission from the hadronic cascade begins to either `cover' the prompt emission spectrum
(see top spectra in Figs.~\ref{sed07} and \ref{sed7}) or peak in the sub-GeV energy band with a 
luminosity approximately equal to that of the MeV emission (see second spectrum from the top in Fig.~\ref{sed07}).
The latter is ruled out by the current status in observations (e.g. \citealt{dermer10}). 
For illustration reasons, the spectra obtained just before
the dominance of the HC are shown with thick lines in Figs.~\ref{sed07} and \ref{sed7}.
From this point on, we will characterize cases that modify the initial MeV emission or have 
$L_{(0.1-1)\rm GeV} \gtrsim 0.2-0.3 \ L_{(0.1-1)\rm MeV}$ commonly
as `HC dominant' cases. We note that if 
a stricter upper limit on the GeV luminosity was used this would also be reflected at the upper limit
imposed on the proton to gamma-ray luminosity ratio discussed in the following section.

\begin{figure}
 \centering
 \includegraphics[width=0.5\textwidth]{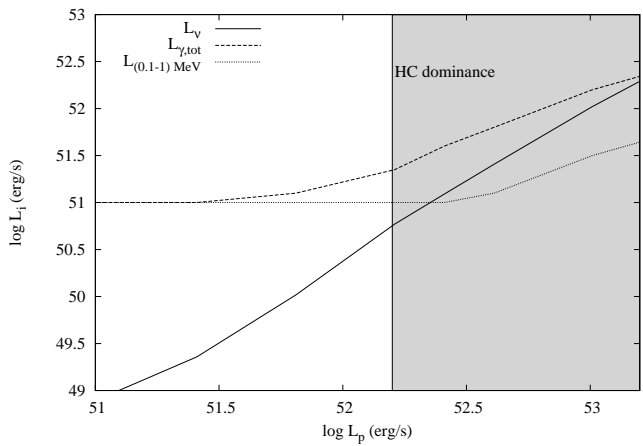}
 \caption{Log-log plot of the total neutrino luminosity (solid line), the bolometric photon luminosity (dashed line) and
 the (0.1-1) MeV gamma-ray luminosity (dotted line) as a function of the proton injection
 luminosity $L_{\rm p}$ for $\ell_{\gamma}=0.7$. The grey colored area is obtained for $L_{\rm p}$ values that lead to the dominance of the 
HC emission.}
 \label{fig2}
\end{figure}

For a fixed $\ell_{\gamma}$, higher secondary emission and  neutrino luminosities can be obtained by increasing
the proton injection compactness.  This is demonstrated in Fig.~\ref{fig2}, where we plot the observed luminosities of 
protons, muon and electron neutrinos,  photons and (0.1-1) MeV gamma-ray photons, for
the case shown in Fig.~\ref{sed07}. For low enough values of $L_{\rm p}$
the bolometric photon luminosity is actual equal to the luminosity emitted in 0.1-1 Mev energy band.
However, for $L_{\rm p}$ above a certain value, which for the particular example is $\simeq 1.5 \times 10^{52}$~erg/s, we obtain 
$L_{\gamma}^{\rm tot} > L_{(0.1-1) \rm MeV}$.
This indicates that the photon component of the hadronic cascade is no longer negligible.
Moreover, we find that $L_{\nu} \propto L_{\rm p}^{\rm q}$, where $q=1$ for
low enough $L_{\rm p}$ but $q=1.6$ for higher proton luminosities. This steeper than linear scaling relation
between $L_{\nu}$ and $L_{\rm p}$ is one more indication of 
the spectral modification due to the HC. Interestingly enough, the 
ratio of the neutrino to the bolometric photon luminosity becomes maximum only for proton luminosities leading to HC dominant cases (grey colored area in Fig.~\ref{fig2}).
Still, for the last case before the dominance of the HC, we  find that 
the photon and neutrino components are energetically equivalent with 
$L_{\nu} \simeq 0.3 L_{\gamma}^{\rm tot}$ (see also Table~\ref{table1}).

\subsection{Maximum proton to gamma-ray luminosity ratio}

For each value of $\ell_{\gamma}$ we can 
derive a maximum value of the proton compactness above which the hadronic cascade significantly alters the GRB photon spectrum, as previously described.
This can also be translated to a maximum value of the ratio $\hp = L_{\rm p} / L_{\gamma}$, which
in terms of compactnesses is written as  $\hp =\lp (\mpr/\mel) / \ell_{\gamma}$\footnote{An alternative definition for $\hp$
would be  $\hp = L_{\rm p} / \Lgtot$, where the superscript `tot' denotes the bolometric photon luminosity. In the present work, however, we adopt 
the definition with the gamma-ray luminosity, as this appears mostly in the literature. }. 
For these maximum values, we then derive (i) the ratio $\eta_{\nu}=L_{\nu}/L_{\gamma}$, (ii) the ratio $\zeta_{\nu} = L_{\nu}/\Lgtot$, (iii)
the production efficiency in neutrinos ($\xi_{\nu}$), and (iv) the production efficiency in 
secondaries that contribute to the HC ($\xi_{\rm sec}$), where the efficiencies are defined
as $\xi_{\nu, \rm sec} = L_{\nu, \rm sec}/ L_{\rm p} = \ell_{\nu, \rm sec} (\mel/\mpr) / \lp$. 
Our results are summarized in Table~\ref{table1} and a few things worth mentioning follow:
\begin{table*}
 \caption{Maximum values of various ratios-efficiencies derived for five values
 of the gamma-ray compactness.}
 \begin{tabular}{ccc  ccc  cc}
 \hline
 $\ell_{\gamma}$ & $\ell_{\gamma}^{\rm tot}$ & $\log\ell_{\rm p, \max}^{\rm inj}$ & $\eta_{\rm p, \max}$ & $\eta_{\nu, \max}$ & $\zeta_{\nu, \max}$& $\xi_{\rm sec, \max}$ ($\%$) & $\xi_{\nu, \max}$ ($\%$)\\
 \hline
 0.07 & 0.2 & $-2.6$ & $ 66$ & 0.9&  0.3  &  $2.8$ & $1.4$\\
 0.22 & 0.9 & $-2.4$ & $35$  & 1.4& 0.35  &  $7$    & $4$  \\
 0.7 & 1.5 & $-2.2$ & $16.5$ &0.6 & 0.28  & $6.7$ & $3.6$ \\
 2.2 & 6.1  & $-1.8 $& $14$ & 1.4 & 0.5  & $15 $   & $10$ \\
 7.0 & 15.7 & $ -1.4$ & $10.4$ & 1.4& 0.6 & $ 17.5$ & $13$ \\
 \hline
 \end{tabular}
\label{table1}
\end{table*}

\begin{figure}
 \centering
 \includegraphics[width=0.5\textwidth]{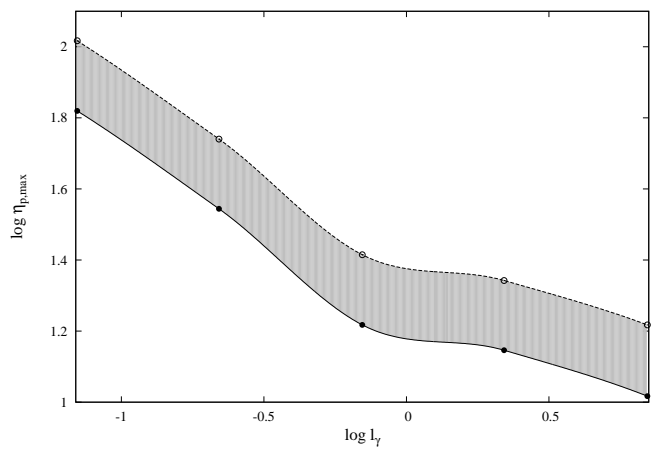}
 \caption{Log-log plot of $\eta_{\rm p, \max}$ as a function of $\ell_{\gamma}$. 
 The actual results of our simulations are shown with points while the lines are the result of interpolation.
 Results shown with filled circles/solid line are obtained for $\eta_{\rm p, \max}$
 while open circles/dashed line correspond to $\hp=\eta_{\rm p, \max}10^{\delta x}$.
 The grey colored area denotes the uncertainty of the derived $\eta_{\rm p, \max}$.
  }
 \label{fig3}
\end{figure}

\begin{itemize}
 \item the maximum value of $\hp$ decreases as the photon compactness becomes larger. This is exemplified in Fig.~\ref{fig3}.
  An extrapolation of our results to even larger values of $\ell_{\gamma}$ (see e.g. \citealt{murase08, zhangkumar13})
  would imply that $L_{\rm p} \approx (1-10) L_{\gamma}$ -- see also \S 5 for a comparison with other works. 
  \item there is an anticorrelation between the maximum allowable value of $\hp$ and $\xi_{\nu}$ ($\zeta_{\nu}$)
  that reflects the fact that the upper limit imposed on $\lp$ by the HC, becomes more stringent as $\ell_{\gamma}$ increases.
   \item the maximum ratio $\zeta_{\nu}$ increases as the gamma-ray compactness becomes larger and may as high as $60 \%$ for $\ell_{\gamma} \gtrsim 7$.
  \item we find that $\eta_{\rm p, \max} \propto 1/\xi_{\rm sec, \max}$, which can be understood as follows the photon compactness from the cascade
  is defined as $\ell_{\rm sec} = \xi_{\rm sec} \lp (\mpr/\mel)$ while $\lp = \hp \ell_{\gamma} (\mel/\mpr)$. Roughly speaking, 
  the secondary photon emission  will start affecting the overall photon spectrum, if $\ell_{\rm sec} \approx \alpha \ell_{\gamma}$ where $\alpha$ is a 
  numerical factor that contains all the details about the hadronic cascade and depends on a series of parameters, such as the magnetic field strength and
  the spectral shape of the secondary emission itself. 
  Our results show, however, that $\alpha$ varies less than a factor of 2 among the five parameter sets presented in Table~\ref{table1}, and we can, therefore, consider it
  a constant. By combining the above we find that $\eta_{\rm p, \max}\approx \alpha/ \xi_{\rm sec, \max} \propto 1/ \xi_{\rm sec, \max} $.  
  \end{itemize}

\subsection{Effects of other parameters}
In the previous section we demonstrated the effects of the HC on the photon and neutrino
 spectra but more importantly we showed that the HC imposes an upper limit on
 the ratio $L_{\rm p}/L_{\gamma}$.  Here, in an attempt to test the robustness 
 of these results, we discuss in detail the effects of the magnetic field strength, of
 the initial spectral shape in gamma-rays,  and of a larger
 value of $\Gamma$.
 \subsubsection{Magnetic field strength}
\begin{figure}
   \centering
    \includegraphics[width=0.5\textwidth]{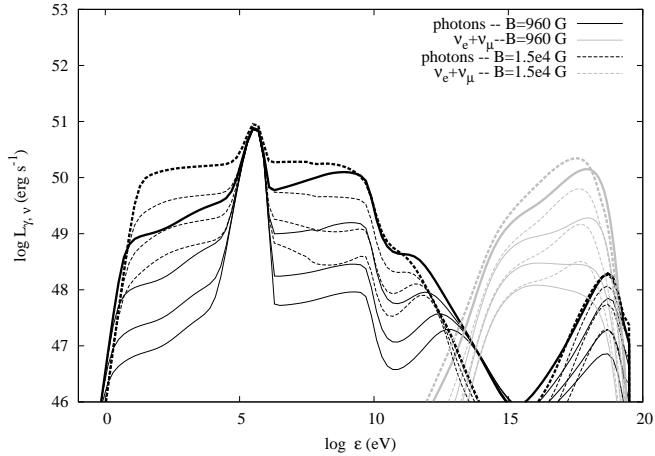}
\caption{Comparison of photon and neutrino spectra for $\ell_{\gamma}=0.7$ and two values of the magnetic field strength, i.e. $B=960$~G and $B=1.5\times10^4$~G, shown
with solid and dashed lines, respectively. The proton injection compactness ranges from $10^{-3.4}$ (bottom) to $10^{-2.2}$ (top). Thick lines
    have the same meaning as in Fig.~\ref{fig1}.}
\label{sed07_Bcomp}
   \end{figure}

In general, stronger magnetic fields enhance the cascade emission in two ways:
\begin{itemize}
 \item through synchrotron radiation of secondary pairs, which leaves an imprint mainly at 
 energies below 1~MeV in the comoving frame because of the strong photon attenuation that
 affects higher energy photons. 
   \item through synchrotron cooling of muons and pions, which starts playing a role for high enough magnetic fields. 
   This introduces more photons into the system, since
   part of the energy that would have been transfered to the neutrino component goes now into the photon component.
\end{itemize}
The effects of the magnetic field on the photon and neutrino emission are exemplified in Fig.~\ref{sed07_Bcomp}.
For the same proton injection compactness, we find that the flux of the HC emission increases, while the neutrino spectra
become harder. The change of the neutrino spectral shape is non-trivial (see also \citealt{asanomeszaros14}) and 
it requires a self-consisent treatment, as it strongly depends on the shape of the target photon field.
Finally, we find that the value of $\eta_{\rm p, \max}$ is not significantly affected by the choice of the magnetic field, while 
the values of $\xi_{\rm sec}$ and $\xi_{\nu}$ 
increase at most by a factor of two -- see Table~\ref{table2}.
Although synchrotron cooling of pions is not relevant
for the benchmark case, it starts playing a role for the higher magnetic field strength considered here.
In particular, we find that 
$E_{\pi, \rm c}\simeq E_{\pi, \max}/10$, where $E_{\pi, \max} \simeq 0.2 \gmx \mpr c^2 = 8\times10^{16}$~eV and
$E_{\pi, \rm c}$ is the typical energy of a pion that cools due to synchrotron radiation before it decays. 
Since the numerical code does not account for the synchrotron losses of pions (see also \S 3), 
the results presented in this section should be considered as an upper limit. Inclusion
of pion cooling 
would have a twofold effect on our results:
first, the peak of the neutrino spectra would move towards smaller energies by approximately an order of magnitude and second,
the resulting deficit in the neutrino luminosity would be balanced accordingly by an increase
of the bolometric photon luminosity. For the top spectrum in Fig.~\ref{sed07_Bcomp} in particular, we estimated
the decrease in the neutrino luminosity after 
inclusion of pion cooling to be $\Delta L_{\nu} \simeq  0.4 \times 10^{51}$~erg/s. This would be translated to a slight
increase of the the bolometric luminosity, i.e. $\Lgtot \sim 4.4\times10^{51}$~erg/s instead of $4\times 10^{51}$~erg/s that
is the value we obtain when we neglect pion synchrotron losses.
Thus, the limits derived for $\hp$ are robust, while the values of $\zeta_{\nu, \max}$ and $\eta_{\nu, \max}$ 
listed in Table~\ref{table2} should be lower by a factor of a few.
   
\begin{figure}
   \centering
    \includegraphics[width=0.5\textwidth]{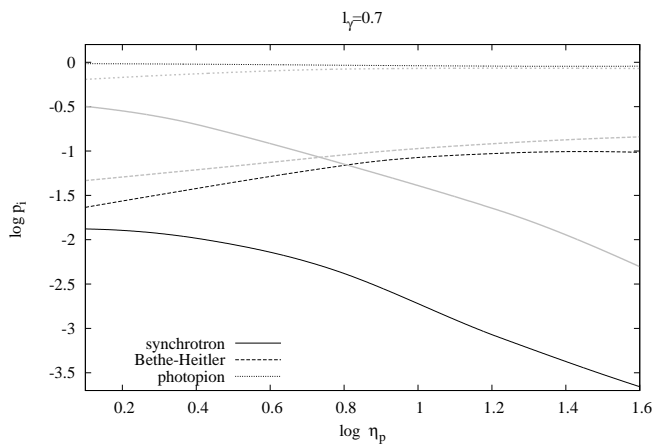}
\caption{Plot of the contribution of synchrotron (solid lines),  photopion (dotted lines) and Bethe-Heitler (dashed lines)
 energy loss rates to the total energy loss rate of protons as a function of the ratio $\hp$ for $\ell_{\gamma}=0.7$ and
two values of the magnetic field strength: $B=960$~G (black lines) and $B=1.5\times10^4$~G (grey lines).  
 Here, $\log \eta_{\rm p}>0.8$ corresponds to HC dominant cases. Thick lines
    have the same meaning as in Fig.~\ref{fig1}. }
\label{losses_comp}
   \end{figure}

It has been already pointed out by various authors (e.g. WB97, \citealt{asano05}), that the photopion processes
is the dominant energy loss mechanism of high-energy protons in sources with large photon compactnesses.
This is also illustrated in Fig.~\ref{losses_comp}, where we plot the 
ratio $p_{\rm i}$ of the proton energy loss rate
due to a process $i$ (synchrotron, Bethe-Heitler and photopion)
to the total energy loss rate. The photopion process dominates indeed,
over all other processes for  the whole range of $\hp$ values. This is also the case for the other values of $\ell_{\gamma}$.
   
\begin{figure}
   \centering
    \includegraphics[width=0.5\textwidth]{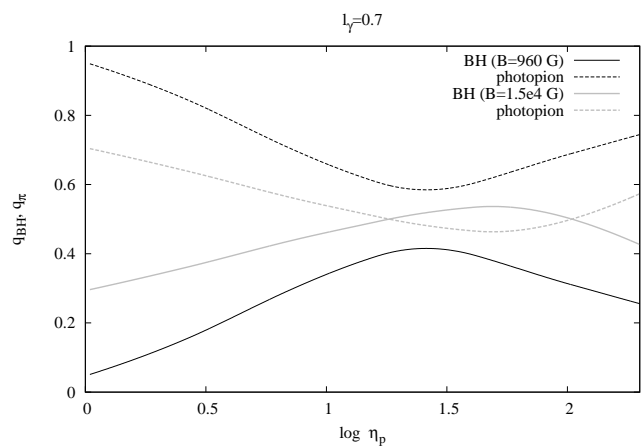}
    \caption{Rate of the energy injection into the hadronic cascade (pairs and photons) by the Bethe-Heitler ($q_{\rm BH}$) and the
    photopion ($q_{\pi}$)
    processes as a function of the ratio $\hp$, for $\ell_{\gamma}=0.7$ and two values of the magnetic field strength: 
    $B=960$~G (black lines) and $B=1.5\times10^4$~G (grey lines). Here, $\log \eta_{\rm p}>0.8$ corresponds to HC dominant cases.}
    \label{injection_comp}
   \end{figure}

The contribution of the Bethe-Heitler process to the hadronic cascade, however, is not always negligible. 
The main reason for this is that only a part of the energy lost by the protons 
through photopion interactions contributes to the hadronic cascade itself; the rest goes to the neutrino and high-energy neutron components.
This is exemplified in Fig.~\ref{injection_comp},  where we plot $q_{\rm BH}$ and $q_{\pi}$ as a function of $\hp$ for $\ell_{\gamma}=0.7$ and two values
of the magnetic field marked on the plot. Here, $q_{\rm BH}$ and $q_{\pi}$ are 
the energy injection rates defined as $q_{\rm BH} = \dot{E}_{\rm BH}/\dot{E}_{t\rm ot}$ and 
$q_{\pi} = (\dot{E}_{p \gamma}^{\pi^{\pm}\rightarrow e^{\pm}} + \dot{E}_{p\gamma}^{\pi^{0}\rightarrow \gamma\gamma})/\dot{E}_{\rm tot}$, where $\dot{E}_{\rm k}$
denotes the energy injection rate of the process $k$ into secondary pairs and/or photons. 
The dependance of the ratios $q_{\rm BH}$ and $q_{\rm p\gamma}$ on $\hp$ is non-trivial and it is mainly
determined by the characteristics of the target photon field, e.g. spectral shape and compactness, which 
highlights once again the role of the HC. From a quantitative point of view, we find that for indermediate values of $\hp$
the contribution of the Bethe-Heitler process to the HC may exceed $50 \%$ and $30 \%$ for 
the high and low magnetic field values, respectively.
\begin{table}
  \caption{Same as in Table~\ref{table1} but for $B=1.5\times 10^4$~G.}
 \begin{tabular}{cccccc}
 \hline
 $\ell_{\gamma}$ & $\eta_{\rm p, \max}$ & $\eta_{\nu, \max}$ & $\zeta_{\nu, \max}$ & $\xi_{\rm sec, \max}$ ($\%$) & $\xi_{\nu, \max}$ ($\%$)\\
 \hline
 0.07 & $66 $ & 1& 0.5 & $7$ & $ 3$\\
 0.7 &  $16.5$ & 1& 0.32 & $14$ & $ 6$ \\
 7 &  $10.4$ & 1.2& 0.35 & $30$ & $ 14 $ \\
 \hline
 \end{tabular}
 \label{table2}
 \end{table}   
 
   \subsubsection{Shape of GRB spectrum}
As stated in \S 2, we adopted a fiducial photon spectrum to describe the GRB prompt emission. We chose, in particular,
a grey-body photon field that can be described only by two parameters and can give us insight on
the effects of the hadronic cascade mainly because of its narrow spectral energy distribution.
In this section we investigate the implications of the above choice on the results presented so far, such as 
the maximum ratio of proton to gamma-ray luminosity. 

The gamma-ray spectrum of the prompt emission can be successfully modelled, at least in most cases\footnote{GRB spectra
that cannot be adequately fitted by the Band model, either show evidence of some black-body component (e.g. \citealt{axelsson12})
or require an additional power-law component (e.g. \citealt{guiriec10}).}, by 
the so-called `Band function' \citep{band93, band09}. 
For the purposes of the present work, it is sufficient to assume a `Band-like' photon spectrum and repeat the proceedure of \S 4.3, without
getting into the details of the emission mechanism itself. We model the gamma-ray photon spectrum as $\nph \propto x^{-\alpha}$, for $x_{\rm min}<x<x_{\rm br}$
and $\nph \propto x^{-\beta}$, for $x_{\rm br}<x<x_{\rm max}$, where $\alpha=1.2$, $\beta=2.2$, $x_{\min}=3\times 10^{-6}$, $x_{\rm br}=3\times 10^{-3}$,
$x_{\max}=0.3$  and $x$ is the photon energy as measured in the comoving frame, in units of $\mel c^2$. 
The gamma-ray photon compactness can be also written as
\eqb
\ell_{\gamma} = \frac{\sth \rb u_{\gamma} }{\mel c^2},
\label{lg-def}
\eqe
where 
\eqb
u_{\gamma} = \frac{L_{\gamma}} { 4 \pi c \rb^2 \Gamma^4}
\label{ug-def}
\eqe
is the energy density of the photon field in the comoving frame. 

\begin{figure}
   \centering
    \includegraphics[width=0.5\textwidth]{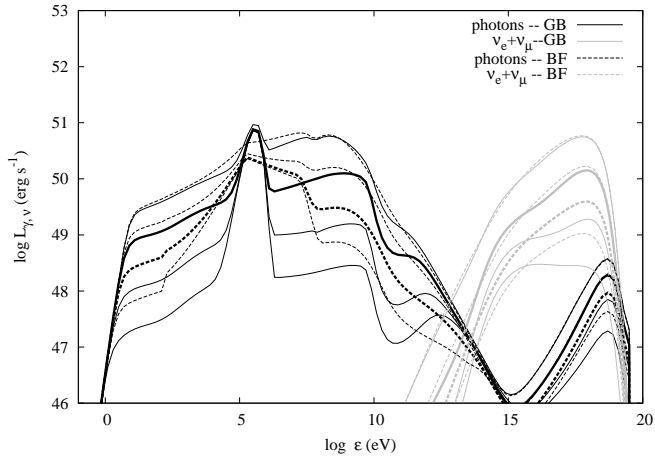}
    \caption{Comparison of photon and neutrino spectra for $\ell_{\gamma}=0.7$ and two 
    different prompt gamma-ray  spectra, i.e.
    a grey-body photon field (GB) and  a Band function spectrum (BF) shown with solid and dashed lines, respectively. The proton injection compactness ranges
    between $10^{-3}$ (bottom curve) and $10^{-1.8}$ (top curve) with logarithmic increaments of 0.4. }
    \label{sed07_Band}
   \end{figure}

A comparison of the multiwavelength photon spectra and neutrino spectra obtained for $\ell_{\gamma}=0.7$ and two different 
initial gamma-ray photon fields is shown in Fig.~\ref{sed07_Band}.
In the case of a Band-like initial photon field, we find that $\eta_{\rm p,\max}=10.4$, i.e. smaller by a factor of 1.6 than the one derived in the case
of a grey-body photon spectrum. This small decrease is found also for other values of the initial photon compactness (see Table~\ref{table3}), and it
can be understood as follows:  in the case of a Band-like photon field, 
the same gamma-ray luminosity or compactness as before is now distributed over a wider energy range. This increases, therefore,
the possible combinations of proton and photon energies with $\gp x$ that lies above the threshold for Bethe-Heitler pair production and/or pion production, and
slightly enhances the emission from the secondaries.
Thus, for the same magnetic field strength, the marginal curve $\eta_{\rm p, \max}\left(\ell_{\gamma}\right)$ derived for a grey-body photon spectrum (see Fig.~\ref{fig3})
should be shifted towards smaller values by a factor of $\sim 1.6$.
   
Furthermore, the values listed in Table~\ref{table3} imply the following rough scaling relations,
\begin{itemize}
 \item$L_{\rm p} \sim 25 L_{\gamma}$, $L_{\nu} \sim 0.5 L_{\gamma}$ and $L_{\nu} \sim 0.2 \Lgtot$,  for $\ell_{\gamma} \ll 1$ \\
 \item  $L_{\rm p} \sim L_{\gamma}$, $L_{\nu} \sim 0.8 L_{\gamma}$ and $L_{\nu} \sim 0.3 \Lgtot$,  for $\ell_{\gamma} \gg 1$,
\end{itemize}
which are robust within a factor of 2 -- see previous section for the effects of the magnetic field strength.
 The values $\ell_{\gamma}=30$ and 70 correspond to observed gamma-ray luminosities $4\times 10^{52}$ erg/s and $10^{53}$ erg/s, respectively. Thus, the derived
values for the various efficiencies are representative for typical and bright GRBs (e.g. \citealt{wandermanpiran10}).
Even if  $\eta_{\rm p, \max} \sim 1-2$ for the brightest GRBs, we still find $L_{\nu}/L_{\gamma} \gtrsim 1$ and thus, these can be still considered significant neutrino emitters.

Besides the marginal effects on the ratios $\hp$ and  $\zeta_{\nu}$, we find that the 
neutrino spectra become harder (see grey lines in Fig.~\ref{sed07_Band}). This, however, 
should not be considered as a general result, since simulations  with Band-like photon spectra and higher $\ell_{\gamma}$ resulted in flat
(in $\nu F_{\nu}$ units) neutrino spectra, i.e. similar to those in Fig.~\ref{sed7}. Our results indicate that
for indermediate photon compactnesses, e.g. $\ell_{\gamma} \sim 1$, the shape of the neutrino spectra is sensitive
to various parameters, such as the magnetic field strength (see Fig.~\ref{sed07_Bcomp}) and the shape
of the initial gamma-ray spectrum, and may deviate from the expected power-law form $\epsilon_{\nu}^{-2}$. 
This requires further investigation which, however, is outside the scope of the present work. 
 
 \begin{table}
 \caption{Same as in Table~\ref{table1} but for a `Band-like' gamma-ray spectrum and 
 two additional values of $\ell_{\gamma}$.}
 \begin{tabular}{cccccc}
 \hline
 $\ell_{\gamma}$ & $\eta_{\rm p, \max}$ & $\eta_{\nu, \max}$& $\zeta_{\nu,\max}$&  $\xi_{\rm sec, \max}$ ($\%$) & $\xi_{\nu, \max}$ ($\%$)\\
 \hline
 0.07 & $ 41.6$ & 0.5& 0.2 & $1.3$ & $0.7$\\
 0.7 &  $10.4$ & 0.5 & 0.2 & $ 6$ & $3$\\
 7 &  $6.6$ & 0.8& 0.5 & $17$ & $13$ \\
 15 &  3 & 0.5 & 0.4& 21 & 17 \\
 30 & 2  &0.8 & 0.3 & 26& 20\\
  70& 1.6  & 1.0  & 0.3    & 53   & 60 \\
 \hline
 \end{tabular}
  \label{table3}
 \end{table}
\subsubsection{Lorentz factor}
For the benchmark case we chose a moderate value of the Lorentz factor (see \S 2). Since there is no reason 
for excluding  values $> 300$  {\sl a priori}, here we examine the implications of a larger $\Gamma$ on our results.
The choice of a larger value of $\Gamma$ has a twofold effect: the peak energy of the GRB spectrum as measured in the comoving frame decreases and
the photon compactness drops significantly, since $\ell_{\gamma} \propto \rb^{-1}\Gamma^{-4}$. We showed
that the maximum baryon loading given by $\eta_{\rm p, \max}$ increases as the emission region becomes
less compact in gamma-rays (see Fig.~\ref{fig3}), whereas the efficiency 
in the production of secondaries, such as pairs and neutrinos, drops (see e.g. Table~\ref{table3}).

As an indicative example, we choose the case with $\ell_{\gamma}=7$ and
a three times larger $\Gamma$ than before, i.e. $\Gamma=708$. The photon compactness now becomes $0.07$ and 
the peak energy of the Band spectrum decreases by a factor of three, i.e. $x_{\rm br}=10^{-3}$.
Following the same steps as in \S 4.2, we find $\eta_{\rm p, \max} =42$. The various efficiencies are  $\eta_{\nu, \max}=0.5$, $\zeta_{\nu, \max}=0.2$,
$\xi_{\rm sec, \max}=0.02$, $\xi_{\nu, \max}=0.02$ and should be compared with the values listed in Table~\ref{table3} for $\ell_{\gamma}=7$.
\begin{figure}
   \centering
    \includegraphics[width=0.5\textwidth]{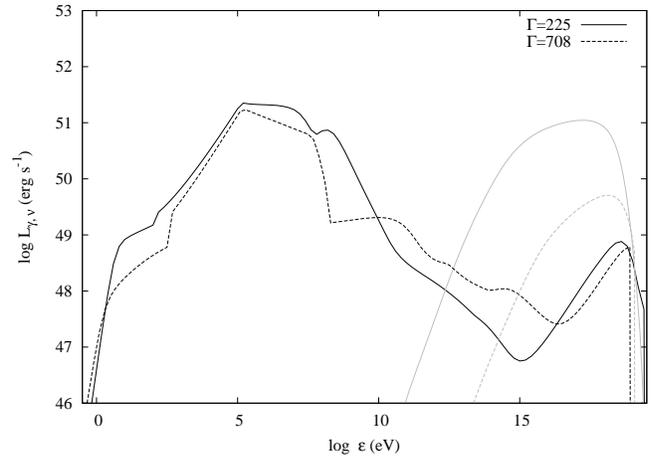}
    \caption{Photon (black lines) and neutrino (grey lines) spectra for $\Gamma=225$ and $\Gamma=708$ shown with solid and dashed lines, respectively. Other
    parameters used are: $\rb=10^{12}$~cm, $B=960$~G, $L_{\gamma}=10^{52}$~erg/s, $L_{\rm p}=6.6 L_{\gamma}$ and $\gmx=4\times 10^8$.}
    \label{fig4}
   \end{figure}   
To illustrate better the above, we compare the photon and neutrino spectra between two cases for $\Gamma=225$ and 708. 
The rest of the parameters, which are kept fixed, are:  $\rb=10^{12}$~cm, $B=960$~G, $L_{\gamma}=10^{52}$~erg/s, $L_{\rm p}=6.6 L_{\gamma}$ and $\gmx=4\times 10^8$.
The photon and neutrino spectra are shown in Fig.~\ref{fig4}. Note that the distortion of the gamma-ray spectrum 
becomes evident only for the first case (black lines), since $\eta_{\rm p}=\eta_{\rm p, \max}=6.6$ (see Table~\ref{table3}).
The main reason for the differences seen in the photon spectra above $\sim 100$~MeV 
and in the neutrino emission is the  photon compactness,  which is  decisive for both the intrinsic 
optical depth for $\gamma \gamma$ absorption and the efficiency of photopion production (see also \citealt{asanoetal09}). 
Finally, the lower peak photon energy in the comoving frame proves to be not as important as the lower value of $\ell_{\gamma}$.

\section{Comparison with other works}
\begin{figure}
   \centering
    \includegraphics[width=0.5\textwidth]{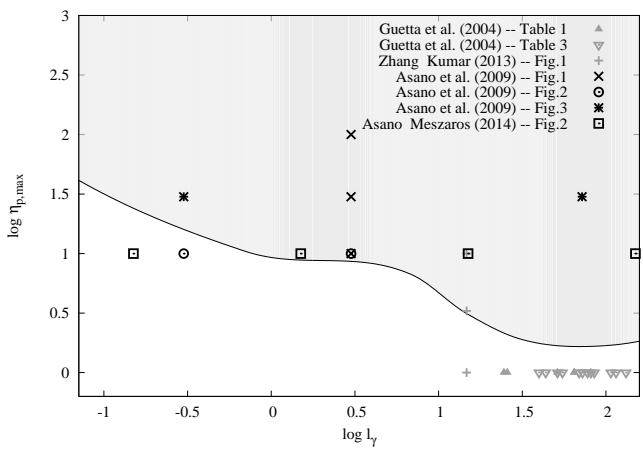}
    \caption{The upper limit $\eta_{\rm p, \max}$ plotted as a function of $\ell_{\gamma}$ (solid line) along with indicative
    values derived from studies that focus on the diffuse GRB neutrino emission (grey symbols)
    and on the HC emission (black symbols). The emission from the HC distorts the GRB spectrum for parameter values
    drawn from the grey-colored region.}
    \label{comp}
   \end{figure}
In this section we attempt a comparison of our results regarding the maximum value
of $\hp$ imposed by the hadronic cascade with other works in the literature.
Our results are summarized in Fig.~\ref{comp}, where the solid line denotes
the maximum value $\eta_{\rm p, \max}$ as a function of the gamma-ray compactness, which
divides the $\hp-\ell_{\gamma}$ into two regions. Above the curve  (grey colored region) the effects of the HC
on the initial gamma-ray spectrum become prominent, whereas 
the combination of $\hp$ and $\ell_{\gamma}$ values drawn from the region 
below the curve does not lead to strong secondary emission.

The steps for deriving $\eta_{\rm p, \max}$ for different
values of $\ell_{\gamma}$ are the same as those described in \S 4.3. 
Here, we assume a Band function spectrum with fixed photon indices
$\alpha, \beta$ and
characteristic energies $x_{\min}, x_{\rm br}$ and $x_{\max}$ as described in \S 4.4, since
this is a common assumption in most studies\footnote{The gamma-ray spectrum in \cite{asanoetal09} is the result
of synchrotron radiation of relativistic electrons, which, however, may be described by a Band function.}.
The magnetic field is taken to be $B=960$~G.  We note also that the limiting curve in  Fig.~\ref{comp} may be shifted for different values
of the spectral indices or/and of the break energy (see also \S 4.4),
but no more than by a factor of 2-3. Thus, 
the upper limit shown in Fig.~\ref{comp} can be considered robust within this factor.

The points shown in Fig.~\ref{comp} correspond to different parameter sets studied in the literature.
In particular, grey colored symbols correspond to studies of the neutrino flux from GRBs, whereas
black colored symbols denote works that focus on the induced hadronic cascades.
A comparison between our upper limit and other works is possible by, first, noting that 
the parameter $\hp=L_{\rm p}/L_{\gamma}$ used here is equivalent to: (i) the parameter
$f_{\gamma/p}^{-1}$ in \cite{zhangkumar13}, (ii) the ratio $\epsilon_{p}/\epsilon_{\rm e}$ in \cite{asanoetal09}, (iii)
the ratio of total energies in protons and gamma-rays in \cite{guetta04}, and (iv) the parameter $f_{\rm p}$ in \cite{asanomeszaros14}.
For the derivation of $\ell_{\gamma}$ (see eq.~(\ref{lg-def})) we combined various parameters given in the aforementioned works, namely
$\Gamma$, $r$, $\dt$, and $L_{\gamma}$ with eq.~(\ref{ug-def}). We note that 
$r=f_{\rm v} c  \Gamma^2 \dt$, where $f_{\rm v}$ is a numerical factor that ranges between 1-2 among the various 
studies. For the compilation of Fig.~\ref{comp}, we used the value of $f_{\rm v}$ as given in each work.  

Let us focus first on the black colored symbols. Most of the points that lie above our upper limit correspond 
to cases with significant modifications of the gamma-ray spectrum
because of the HC in agreement with the respective studies. For example, the first two curves from top of Fig.~3  in \cite{asanoetal09}
are indicative cases of proton-cascade dominated spectra. 
Using the same values as in \cite{asanoetal09}, we find that the gamma-ray compactness is $\ell_{\gamma}=0.3$ and 72 for $\Gamma=300$ and 100, respectively. 
These are shown as black stars in Fig.~\ref{comp}.
Given that the depicted values are obtained for a wide range of physical parameters, such as $\Gamma$, $L_{\gamma}$ and 
$\rb$, the plane $\hp-\ell_{\gamma}$ proves to be a robust tool for distinguishing between cases with 
dominant hadronic cascade or not. For the former, the neutrino and secondary production efficiencies 
may exceed the maximum values derived in previous sections (see also Fig.~\ref{fig2}).
GeV bright GRBs detected by {\sl Fermi}-LAT  are a good example of bursts that deviate from  the `typical' ones (e.g. \citealt{racusin08,abdo09}). 
Assuming that their high-energy emission is a result of the hadronic initiated cascade (e.g. \citealt{asanoinoue10}), these bursts might 
fall into the grey-colored region of Fig.~\ref{comp} depending on the relative ratio of the GeV/MeV fluxes: 
bright bursts with comparable fluxes in the two energy bands would be also a good candidate source of bright neutrino emission.

The grey colored symbols lie, in general, at the right part of the $\hp-\ell_{\gamma}$ plane,
since high gamma-ray compactnesses are required for efficient neutrino production.  Interestingly, most of these points
lie below but close (within a factor of two) to our limiting curve.
For example, if a similar analysis to the one by \cite{guetta04} was performed, 
but for a higher ratio of proton to gamma-ray luminosities, e.g. $\eta_p > 3$, 
the distortion of the {\sl pre-assumed} GRB spectrum could not be avoided. Thus, in the regime of large 
gamma-ray compactnesses the calculation of the neutrino emission necessitates the treatment of the secondary photon emission as well.

Note that the above discussion does not necessarily refer to bright GRBs. Because of the strong dependance of the gamma-ray compactness
to the Lorentz factor, namely $\ell_{\gamma} \propto L_{\gamma} / \Gamma^5 \dt$, a typical burst with $L_{\gamma}=3\times 10^{52}$~erg/s
may have ten times larger $\ell_{\gamma}$ than a bright GRB with $L_{\gamma}=10^{54}$~erg/s, if its bulk Lorentz factor is by a factor of three smaller. 
It is noteworthy that the same conclusion is reached using different argumentation by \cite{asanomeszaros14}. 
Although $L_{\gamma}$ is relatively constrained and it lies in the range $10^{51}-10^{53}$~erg/s, there is freedom 
in the choice of $\Gamma$ and $\dt$, which makes impossible the definition of an observationally typical $\ell_{\gamma}$. Thus,
studying the effects of the hadronic cascade using as a free parameter the gamma-ray compactness covers a wide range
of GRBs and may  facilitate future parameter studies of GRB neutrino emission.


\section{Summary and discussion}
We investigated the effects of hadronic cascades on the gamma-ray burst (GRB) prompt emission spectra
in scenarios of efficient neutrino production.
For this, we employed a generic method where we approximated the prompt GRB emission by either
a grey-body or a Band-function spectrum, while we used as free parameters the proton injection and the prompt gamma-ray 
compactnesses. Using a numerical code that follows  the 
evolution of the proton, photon, neutrino and pair distributions both in energy and time, we 
calculated the steady-state photon and neutrino emission spectra taking into account in a self-consisent way the formation
of the hadronic cascade. We showed that for each value of the  gamma-ray compactness one can set an upper
limit to the ratio of the proton to gamma-ray luminosity by using the fact that the emission from 
the hadronic cascade cannot always be neglected. On the contrary, it is an important
ingredient of GRB hadronic models and it may significantly affect the overall photon 
and neutrino emission.

The recent PeV neutrino detection by the  IceCube Collaboration 
\citep{icecube13} is beginning to place stonger constraints on the parameter values used in various GRB models \citep{zhangkumar13}.
Using a detailed example (see Fig.~\ref{fig2}) we pointed out
that parameter sets that lead to high neutrino luminosities ($ \eta_{\nu} =L_{\nu}/L_{\gamma} \gg 1$) also result 
in modified gamma-ray photon spectra due to the emission from the 
hadronic cascade. On the other hand, we showed that the ratio $\eta_{\nu, \max}$, which is derived for parameter values just before the dominance of the hadronic cascade,
lies in the range $0.5-1.4$. Higher values are obtained for stronger  magnetic fields (see Table~\ref{table2}) and for 
harder/softer GRB spectra below/above the peak,  i.e. for spectra more similar to a grey-body photon emission (see Tables~\ref{table1} and \ref{table3}).
Using this upper limit on the ratio $L_{\nu}/L_{\gamma}$ from a single burst, we can make
a rough estimate of the neutrino energy flux as follows. First, we use  $\bar{\eta}_{\nu} = 0.8$ as an indicative value.
Then,  for a burst with $L_{\gamma}=10^{52}$~erg/s we adopt $E_{\gamma}=4\times 10^{52}$~erg as the typical energy emitted in gamma-rays \citep{ghirlanda12, kakuwa12}.
For the GRB rate at the local universe we use $R_{\rm GRB} \simeq 1$~Gpc$^{-3}$ yr$^{-1}$ \citep{wandermanpiran10}.
We find that the energy production rate of gamma-rays per unit volume  is $ \dot{U}_{\gamma} \simeq 4 \times 10^{43} E_{\gamma, 52.6}$~erg~Mpc$^{-3}$~yr$^{-1}$.
Then, the present day all-flavour neutrino energy density is given by 
$U_{\nu} \approx \bar{\eta}_{\nu} \dot{U}_{\gamma} T$, where $T\simeq 10^{10}$ yr is the Hubble time. Finally, the neutrino flux
is given by
\eqb
E_{\nu}^2\Phi_{\nu} \approx \frac{c}{4 \pi} U_{\nu} \approx  10^{-8} \frac{\bar{\eta}_{\nu}}{0.8} E_{\gamma, 52.6} \ {\rm GeV cm}^{-2} {\rm s}^{-1} {\rm sr}^{-1}.
\label{estimate}
\eqe
The above rough estimate is  lower only by a factor of $\sim 3$ than 
the WB upper bound \citep{wb99} and the neutrino excess measured in the IceCube signal \citep{icecube13} and, interestingly enough, this 
was obtained using an indirect method, i.e. using arguments about the HC dominance on the GRB photon spectra. 
 An implicit assumption for the above derivation is that typical GRBs are the source of the PeV neutrino emission observed
with IceCube. The non-detection of neutrinos from {\sl Fermi} triggered GRBs, however, 
starts to challenge this assumption (e.g. \citealt{abbasi12, heetal12,liuwang13}). Even if future observations work against this scenario, models
where low-luminosity GRBs ($\sim 10^{47}$~erg/s) are the main contributors to the diffuse PeV emission 
cannot be ruled out, as long as the statistics of their population remains undetermined (see e.g. \citealt{cholishooper13, murase13} for relevant discussion). 
Since the analysis presented here is not based on the absolute value of the GRB luminosity but on parameters that characterize the intrinsic properties
of the emission region, one can repeat the above calculation using the appropriate values for $E_{\gamma}$, $R_{\rm GRB}$ and $\bar{\eta}_{\nu}$.

Our results also indicate that a higher value for the neutrino flux is possible if $\eta_{\nu, \max} \gtrsim 1.3$,
which further implies that if the observed neutrino signal has indeed a GRB origin then  typical
GRB sources should be strongly magnetized with $B$ above a few kG (see \S 4.4.1) and with significant contribution 
of the hadronic cascade to the overall emission. 
Going one step further, one can argue that diffuse neutrino emission models that 
do no follow in detail the formation of the hadronic cascade may give
fluxes  erroneously  close to or even above the WB upper bound simply because 
their parameters are pushed into the HC dominance regime.
 An additional feature of the high-energy neutrino spectrum in the
case of a strong magnetic field would be a cutoff at a a few PeV because of strong synchrotron
cooling of pions and muons. It can be shown
that for $B=10^5$~G, $r=10^{13}$~cm and  $\Gamma$=300 
the spectrum of muons produced via pion decay 
has a cutoff at $\sim 10$~PeV (e.g. \citealt{heetal12}).

The shape of the neutrino spectra is strongly affected by the photon spectrum that
serves as a target field for photopion interactions and, hence, for neutrino production. 
Although the emitted neutrino spectra
are expected to be flat in $\nu F_{\nu}$ units whenever the cooling of a $\gp^{-2}$ proton distribution is efficient and 
the target spectra are `Band-like', this is not always the case. 
Even in the simplest scenario where the only available target photon field is the Band-like
gamma-ray emission, the neutrino spectral
shape also depends on 
the particular pion production channel, e.g. $\Delta^{+}$ resonance or  multipion production channel,  
that is responsible for the neutrino emission at the specific energy band \citep{baerwald11}. To make things even more complicated, photons produced by neutral pion decay or emitted  
by secondary relativistic pairs 
through synchrotron and inverse Compton scattering contribute to the overall multiwavelength emission
and may serve as targets
for photopion interactions, too. The spectral shape of the hadronic cascade emission is, therefore, 
an important factor for the determination of the final neutrino spectral shape.

Furthermore, our analysis showed that the investigation of the available parameter space in GRB models
may be significantly simplified if this is performed using as basic variables the proton and gamma-ray compactnesses instead 
of the typical GRB model parameters, such as the bulk Lorentz factor and the observed isotropic gamma-ray luminosity.
After all, 
it is the compactness of the emitting region in terms of relativistic protons and photons that
mainly determines the efficiencies of  the neutrino and secondary particle production.

Finally, we demonstrated through indicative examples the role of the Bethe-Heitler process in the formation of the hadronic cascade.
This is not always negligible, as its contribution to the injection of secondary pairs into the hadronic cascade
may exceed 30-50$\%$ for $\ell_{\gamma} \sim 0.01-1$, while stronger magnetic fields tend to enhance the role
of the Bethe-Heilter process. However, we find that its contribution is significantly suppressed for $\ell_{\gamma} \gg 10$.

We have explored the role of hadronic cascades in GRB models of efficient neutrino production using
a method that is sufficiently generic to be applicable to different scenarios of prompt GRB emission.
Assuming that GRBs are the sources of the TeV-PeV neutrinos detected by IceCube, 
then our results suggest that GRBs can account for the 
current neutrino flux level only
if there is a substantial contribution of the hadronic emission to the overall
GRB photon spectra.

\section*{Acknowledgements}
I would like to thank 
Dr. S. Dimitrakoudis for providing the numerical code and
Professors A. Mastichiadis and D. Giannios for 
fruitful discussions and comments on the manuscript.
I would also like to thank the anonymous referee for
 his/her comments that helped to improve the manuscript.
Support for this work was provided by NASA 
through Einstein Postdoctoral 
Fellowship grant number PF 140113 awarded by the Chandra X-ray 
Center, which is operated by the Smithsonian Astrophysical Observatory
for NASA under contract NAS8-03060.
\bibliographystyle{mn2e} 
\bibliography{grbhadro} 

\begin{thebibliography}{51}
\expandafter\ifx\csname natexlab\endcsname\relax\def\natexlab#1{#1}\fi

\bibitem[{{Aartsen} {et~al}\mbox{.}(2014){Aartsen}, {Ackermann}, {Adams},
  {Aguilar}, {Ahlers}, {Ahrens}, {Altmann}, {Anderson}, {Arguelles}, {Arlen},
  \& et~al.}]{aartsen14}
{Aartsen} M.~G. {et~al.}, 2014, ArXiv e-prints

\bibitem[{{Abbasi} {et~al}\mbox{.}(2012){Abbasi}, {Abdou}, {Abu-Zayyad},
  {Ackermann}, {Adams}, {Aguilar}, {Ahlers}, {Altmann}, {Andeen}, {Auffenberg},
  \& et~al.}]{abbasi12}
{Abbasi} R. {et~al.}, 2012, Nature, 484, 351

\bibitem[{{Abbasi} {et~al}\mbox{.}(2010){Abbasi}, {Abdou}, {Abu-Zayyad},
  {Adams}, {Aguilar}, {Ahlers}, {Andeen}, {Auffenberg}, {Bai}, {Baker}, \&
  et~al.}]{abbasi10}
{Abbasi} R. {et~al.}, 2010, ApJ, 710, 346

\bibitem[{{Abdo} {et~al}\mbox{.}(2009){Abdo}, {Ackermann}, {Ajello}, {Asano},
  {Atwood}, {Axelsson}, {Baldini}, {Ballet}, {Barbiellini}, {Baring},
  {Bastieri}, {Bechtol}, {Bellazzini}, {Berenji}, {Bhat}, {Bissaldi},
  {Blandford}, {Bloom}, {Bonamente}, {Borgland}, {Bouvier}, {Bregeon}, {Brez},
  {Briggs}, {Brigida}, {Bruel}, {Burgess}, {Burrows}, {Buson}, {Caliandro},
  {Cameron}, {Caraveo}, {Casandjian}, {Cecchi}, {{\c C}elik}, {Chekhtman},
  {Cheung}, {Chiang}, {Ciprini}, {Claus}, {Cohen-Tanugi}, {Cominsky},
  {Connaughton}, {Conrad}, {Cutini}, {d'Elia}, {Dermer}, {de Angelis}, {de
  Palma}, {Digel}, {Dingus}, {Silva}, {Drell}, {Dubois}, {Dumora}, {Farnier},
  {Favuzzi}, {Fegan}, {Finke}, {Fishman}, {Focke}, {Fortin}, {Frailis},
  {Fukazawa}, {Funk}, {Fusco}, {Gargano}, {Gehrels}, {Germani}, {Giavitto},
  {Giebels}, {Giglietto}, {Giordano}, {Glanzman}, {Godfrey}, {Goldstein},
  {Granot}, {Greiner}, {Grenier}, {Grove}, {Guillemot}, {Guiriec}, {Hanabata},
  {Harding}, {Hayashida}, {Hays}, {Horan}, {Hughes}, {Jackson},
  {J{\'o}hannesson}, {Johnson}, {Johnson}, {Johnson}, {Kamae}, {Katagiri},
  {Kataoka}, {Kawai}, {Kerr}, {Kippen}, {Kn{\"o}dlseder}, {Kocevski}, {Komin},
  {Kouveliotou}, {Kuss}, {Lande}, {Latronico}, {Lemoine-Goumard}, {Longo},
  {Loparco}, {Lott}, {Lovellette}, {Lubrano}, {Madejski}, {Makeev},
  {Mazziotta}, {McBreen}, {McEnery}, {McGlynn}, {Meegan}, {M{\'e}sz{\'a}ros},
  {Meurer}, {Michelson}, {Mitthumsiri}, {Mizuno}, {Moiseev}, {Monte},
  {Monzani}, {Moretti}, {Morselli}, {Moskalenko}, {Murgia}, {Nakamori},
  {Nolan}, {Norris}, {Nuss}, {Ohno}, {Ohsugi}, {Omodei}, {Orlando}, {Ormes},
  {Paciesas}, {Paneque}, {Panetta}, {Pelassa}, {Pepe}, {Pesce-Rollins},
  {Petrosian}, {Piron}, {Porter}, {Preece}, {Rain{\`o}}, {Rando}, {Rau},
  {Razzano}, {Razzaque}, {Reimer}, {Reimer}, {Reposeur}, {Ritz}, {Rochester},
  {Rodriguez}, {Roming}, {Roth}, {Ryde}, {Sadrozinski}, {Sanchez}, {Sander},
  {Saz Parkinson}, {Scargle}, {Schalk}, {Sgr{\`o}}, {Siskind}, {Smith},
  {Spinelli}, {Stamatikos}, {Stecker}, {Stratta}, {Strickman}, {Suson},
  {Swenson}, {Tajima}, {Takahashi}, {Tanaka}, {Thayer}, {Thayer}, {Thompson},
  {Tibaldo}, {Torres}, {Tosti}, {Tramacere}, {Uchiyama}, {Uehara}, {Usher},
  {van der Horst}, {Vasileiou}, {Vilchez}, {Vitale}, {von Kienlin}, {Waite},
  {Wang}, {Wilson-Hodge}, {Winer}, {Wood}, {Yamazaki}, {Ylinen}, \&
  {Ziegler}}]{abdo09}
{Abdo} A.~A. {et~al.}, 2009, ApJL, 706, L138

\bibitem[{{Aharonian}(2000)}]{aharonian00}
{Aharonian} F.~A., 2000, New Astronomy, 5, 377

\bibitem[{{Asano}(2005)}]{asano05}
{Asano} K., 2005, ApJ, 623, 967

\bibitem[{{Asano} {et~al}\mbox{.}(2009){Asano}, {Guiriec}, \&
  {M{\'e}sz{\'a}ros}}]{asanoetal09}
{Asano} K., {Guiriec} S., {M{\'e}sz{\'a}ros} P., 2009, ApJL, 705, L191

\bibitem[{{Asano} \& {Inoue}(2007)}]{asanoinoue07}
{Asano} K., {Inoue} S., 2007, ApJ, 671, 645

\bibitem[{{Asano} {et~al}\mbox{.}(2010){Asano}, {Inoue}, \&
  {M{\'e}sz{\'a}ros}}]{asanoinoue10}
{Asano} K., {Inoue} S., {M{\'e}sz{\'a}ros} P., 2010, ApJL, 725, L121

\bibitem[{{Asano} \& {Meszaros}(2014)}]{asanomeszaros14}
{Asano} K., {Meszaros} P., 2014, ArXiv e-prints

\bibitem[{{Axelsson} {et~al}\mbox{.}(2012){Axelsson}, {Baldini}, {Barbiellini},
  {Baring}, {Bellazzini}, {Bregeon}, {Brigida}, {Bruel}, {Buehler},
  {Caliandro}, {Cameron}, {Caraveo}, {Cecchi}, {Chaves}, {Chekhtman}, {Chiang},
  {Claus}, {Conrad}, {Cutini}, {D'Ammando}, {de Palma}, {Dermer}, {Silva},
  {Drell}, {Favuzzi}, {Fegan}, {Ferrara}, {Focke}, {Fukazawa}, {Fusco},
  {Gargano}, {Gasparrini}, {Gehrels}, {Germani}, {Giglietto}, {Giroletti},
  {Godfrey}, {Guiriec}, {Hadasch}, {Hanabata}, {Hayashida}, {Hou}, {Iyyani},
  {Jackson}, {Kocevski}, {Kuss}, {Larsson}, {Larsson}, {Longo}, {Loparco},
  {Lundman}, {Mazziotta}, {McEnery}, {Mizuno}, {Monzani}, {Moretti},
  {Morselli}, {Murgia}, {Nuss}, {Nymark}, {Ohno}, {Omodei}, {Pesce-Rollins},
  {Piron}, {Pivato}, {Racusin}, {Rain{\`o}}, {Razzano}, {Razzaque}, {Reimer},
  {Roth}, {Ryde}, {Sanchez}, {Sgr{\`o}}, {Siskind}, {Spandre}, {Spinelli},
  {Stamatikos}, {Tibaldo}, {Tinivella}, {Usher}, {Vandenbroucke}, {Vasileiou},
  {Vianello}, {Vitale}, {Waite}, {Winer}, {Wood}, {Burgess}, {Bhat},
  {Bissaldi}, {Briggs}, {Connaughton}, {Fishman}, {Fitzpatrick}, {Foley},
  {Gruber}, {Kippen}, {Kouveliotou}, {Jenke}, {McBreen}, {McGlynn}, {Meegan},
  {Paciesas}, {Pelassa}, {Preece}, {Tierney}, {von Kienlin}, {Wilson-Hodge},
  {Xiong}, \& {Pe'er}}]{axelsson12}
{Axelsson} M. {et~al.}, 2012, ApJL, 757, L31

\bibitem[{{Baerwald} {et~al}\mbox{.}(2013){Baerwald}, {Bustamante}, \&
  {Winter}}]{baerwald13}
{Baerwald} P., {Bustamante} M., {Winter} W., 2013, ApJ, 768, 186

\bibitem[{{Baerwald} {et~al}\mbox{.}(2011){Baerwald}, {H{\"u}mmer}, \&
  {Winter}}]{baerwald11}
{Baerwald} P., {H{\"u}mmer} S., {Winter} W., 2011, PhysRevD, 83, 067303

\bibitem[{{Band} {et~al}\mbox{.}(1993){Band}, {Matteson}, {Ford}, {Schaefer},
  {Palmer}, {Teegarden}, {Cline}, {Briggs}, {Paciesas}, {Pendleton}, {Fishman},
  {Kouveliotou}, {Meegan}, {Wilson}, \& {Lestrade}}]{band93}
{Band} D. {et~al.}, 1993, ApJ, 413, 281

\bibitem[{{Band} {et~al}\mbox{.}(2009){Band}, {Axelsson}, {Baldini},
  {Barbiellini}, {Baring}, {Bastieri}, {Battelino}, {Bellazzini}, {Bissaldi},
  {Bogaert}, {Bonnell}, {Chiang}, {Cohen-Tanugi}, {Connaughton}, {Cutini}, {de
  Palma}, {Dingus}, {do Couto e Silva}, {Fishman}, {Galli}, {Gehrels},
  {Giglietto}, {Granot}, {Guiriec}, {Hughes}, {Kamae}, {Komin}, {Kuehn},
  {Kuss}, {Longo}, {Lubrano}, {Kippen}, {Mazziotta}, {McEnery}, {McGlynn},
  {Moretti}, {Nakamori}, {Norris}, {Ohno}, {Olivo}, {Omodei}, {Pelassa},
  {Piron}, {Preece}, {Razzano}, {Russell}, {Ryde}, {Saz Parkinson}, {Scargle},
  {Sgr{\`o}}, {Shimokawabe}, {Smith}, {Spandre}, {Spinelli}, {Stamatikos},
  {Winer}, \& {Yamazaki}}]{band09}
{Band} D.~L. {et~al.}, 2009, ApJ, 701, 1673

\bibitem[{{B{\"o}ttcher} \& {Dermer}(1998)}]{boettcherdermer98}
{B{\"o}ttcher} M., {Dermer} C.~D., 1998, ApJL, 499, L131

\bibitem[{{Cholis} \& {Hooper}(2013)}]{cholishooper13}
{Cholis} I., {Hooper} D., 2013, JCAP, 6, 30

\bibitem[{{Coppi} \& {Blandford}(1990)}]{coppiblandford90}
{Coppi} P.~S., {Blandford} R.~D., 1990, MNRAS, 245, 453

\bibitem[{{Dermer}(2010)}]{dermer10}
{Dermer} C.~D., 2010, in American Institute of Physics Conference Series, Vol.
  1279, American Institute of Physics Conference Series, {Kawai} N., {Nagataki}
  S., eds., pp. 191--199

\bibitem[{{Dermer} \& {Atoyan}(2003)}]{dermeratoyan03}
{Dermer} C.~D., {Atoyan} A., 2003, Physical Review Letters, 91, 071102

\bibitem[{{Dimitrakoudis} {et~al}\mbox{.}(2012){Dimitrakoudis}, {Mastichiadis},
  {Protheroe}, \& {Reimer}}]{dmpr12}
{Dimitrakoudis} S., {Mastichiadis} A., {Protheroe} R.~J., {Reimer} A., 2012,
  A\&A, 546, A120

\bibitem[{{Dimitrakoudis} {et~al}\mbox{.}(2014){Dimitrakoudis}, {Petropoulou},
  \& {Mastichiadis}}]{dpm14}
{Dimitrakoudis} S., {Petropoulou} M., {Mastichiadis} A., 2014, Astroparticle
  Physics, 54, 61

\bibitem[{{Ghirlanda} {et~al}\mbox{.}(2012){Ghirlanda}, {Nava}, {Ghisellini},
  {Celotti}, {Burlon}, {Covino}, \& {Melandri}}]{ghirlanda12}
{Ghirlanda} G., {Nava} L., {Ghisellini} G., {Celotti} A., {Burlon} D., {Covino}
  S., {Melandri} A., 2012, MNRAS, 420, 483

\bibitem[{{Guetta} {et~al}\mbox{.}(2004){Guetta}, {Hooper},
  {Alvarez-Mun\~{}Iz}, {Halzen}, \& {Reuveni}}]{guetta04}
{Guetta} D., {Hooper} D., {Alvarez-Mun\~{}Iz} J., {Halzen} F., {Reuveni} E.,
  2004, Astroparticle Physics, 20, 429

\bibitem[{{Guiriec} {et~al}\mbox{.}(2010){Guiriec}, {Briggs}, {Connaugthon},
  {Kara}, {Daigne}, {Kouveliotou}, {van der Horst}, {Paciesas}, {Meegan},
  {Bhat}, {Foley}, {Bissaldi}, {Burgess}, {Chaplin}, {Diehl}, {Fishman},
  {Gibby}, {Giles}, {Goldstein}, {Greiner}, {Gruber}, {von Kienlin}, {Kippen},
  {McBreen}, {Preece}, {Rau}, {Tierney}, \& {Wilson-Hodge}}]{guiriec10}
{Guiriec} S. {et~al.}, 2010, ApJ, 725, 225

\bibitem[{{He} {et~al}\mbox{.}(2012){He}, {Liu}, {Wang}, {Nagataki}, {Murase},
  \& {Dai}}]{heetal12}
{He} H.-N., {Liu} R.-Y., {Wang} X.-Y., {Nagataki} S., {Murase} K., {Dai} Z.-G.,
  2012, ApJ, 752, 29

\bibitem[{{Hillas}(1984)}]{hillas84}
{Hillas} A.~M., 1984, ARA\&A, 22, 425

\bibitem[{{IceCube Collaboration}(2013)}]{icecube13}
{IceCube Collaboration}, 2013, Science, 342

\bibitem[{{Kakuwa} {et~al}\mbox{.}(2012){Kakuwa}, {Murase}, {Toma}, {Inoue},
  {Yamazaki}, \& {Ioka}}]{kakuwa12}
{Kakuwa} J., {Murase} K., {Toma} K., {Inoue} S., {Yamazaki} R., {Ioka} K.,
  2012, MNRAS, 425, 514

\bibitem[{{Lightman} \& {Zdziarski}(1987)}]{lightmanzd87}
{Lightman} A.~P., {Zdziarski} A.~A., 1987, ApJ, 319, 643

\bibitem[{{Liu} \& {Wang}(2013)}]{liuwang13}
{Liu} R.-Y., {Wang} X.-Y., 2013, ApJ, 766, 73

\bibitem[{{Mannheim} {et~al}\mbox{.}(2001){Mannheim}, {Protheroe}, \&
  {Rachen}}]{mannheim01}
{Mannheim} K., {Protheroe} R.~J., {Rachen} J.~P., 2001, PhysRevD, 63, 023003

\bibitem[{{Mastichiadis} \& {Kirk}(1995)}]{mastkirk95}
{Mastichiadis} A., {Kirk} J.~G., 1995, A\&A, 295, 613

\bibitem[{{Mastichiadis} {et~al}\mbox{.}(2005){Mastichiadis}, {Protheroe}, \&
  {Kirk}}]{mastetal05}
{Mastichiadis} A., {Protheroe} R.~J., {Kirk} J.~G., 2005, A\&A, 433, 765

\bibitem[{{M{\"u}cke} {et~al}\mbox{.}(2000){M{\"u}cke}, {Engel}, {Rachen},
  {Protheroe}, \& {Stanev}}]{muecke00}
{M{\"u}cke} A., {Engel} R., {Rachen} J.~P., {Protheroe} R.~J., {Stanev} T.,
  2000, Computer Physics Communications, 124, 290

\bibitem[{{Murase}(2008)}]{murase08}
{Murase} K., 2008, PhysRevD, 78, 101302

\bibitem[{{Murase} \& {Ioka}(2013)}]{murase13}
{Murase} K., {Ioka} K., 2013, Physical Review Letters, 111, 121102

\bibitem[{{Paczynski} \& {Xu}(1994)}]{paczynskixu94}
{Paczynski} B., {Xu} G., 1994, ApJ, 427, 708

\bibitem[{{Petropoulou} \& {Mastichiadis}(2012)}]{petromast12b}
{Petropoulou} M., {Mastichiadis} A., 2012, MNRAS, 421, 2325

\bibitem[{{Piran}(2004)}]{piran04}
{Piran} T., 2004, Reviews of Modern Physics, 76, 1143

\bibitem[{{Protheroe} \& {Johnson}(1996)}]{protheroe96}
{Protheroe} R.~J., {Johnson} P.~A., 1996, Astroparticle Physics, 4, 253

\bibitem[{{Rachen} \& {M{\'e}sz{\'a}ros}(1998)}]{rachenmeszaros98}
{Rachen} J.~P., {M{\'e}sz{\'a}ros} P., 1998, PhysRevD, 58, 123005

\bibitem[{{Racusin} {et~al}\mbox{.}(2008){Racusin}, {Karpov}, {Sokolowski},
  {Granot}, {Wu}, {Pal'Shin}, {Covino}, {van der Horst}, {Oates}, {Schady},
  {Smith}, {Cummings}, {Starling}, {Piotrowski}, {Zhang}, {Evans}, {Holland},
  {Malek}, {Page}, {Vetere}, {Margutti}, {Guidorzi}, {Kamble}, {Curran},
  {Beardmore}, {Kouveliotou}, {Mankiewicz}, {Melandri}, {O'Brien}, {Page},
  {Piran}, {Tanvir}, {Wrochna}, {Aptekar}, {Barthelmy}, {Bartolini}, {Beskin},
  {Bondar}, {Bremer}, {Campana}, {Castro-Tirado}, {Cucchiara}, {Cwiok},
  {D'Avanzo}, {D'Elia}, {Della Valle}, {de Ugarte Postigo}, {Dominik},
  {Falcone}, {Fiore}, {Fox}, {Frederiks}, {Fruchter}, {Fugazza}, {Garrett},
  {Gehrels}, {Golenetskii}, {Gomboc}, {Gorosabel}, {Greco}, {Guarnieri},
  {Immler}, {Jelinek}, {Kasprowicz}, {La Parola}, {Levan}, {Mangano}, {Mazets},
  {Molinari}, {Moretti}, {Nawrocki}, {Oleynik}, {Osborne}, {Pagani}, {Pandey},
  {Paragi}, {Perri}, {Piccioni}, {Ramirez-Ruiz}, {Roming}, {Steele}, {Strom},
  {Testa}, {Tosti}, {Ulanov}, {Wiersema}, {Wijers}, {Winters}, {Zarnecki},
  {Zerbi}, {M{\'e}sz{\'a}ros}, {Chincarini}, \& {Burrows}}]{racusin08}
{Racusin} J.~L. {et~al.}, 2008, Nature, 455, 183

\bibitem[{{Reynoso}(2014)}]{reynoso14}
{Reynoso} M.~M., 2014, ArXiv e-prints

\bibitem[{{Vietri}(1995)}]{vietri95}
{Vietri} M., 1995, ApJ, 453, 883

\bibitem[{{Wanderman} \& {Piran}(2010)}]{wandermanpiran10}
{Wanderman} D., {Piran} T., 2010, MNRAS, 406, 1944

\bibitem[{{Waxman}(1995)}]{waxman95}
{Waxman} E., 1995, ApJL, 452, L1

\bibitem[{{Waxman} \& {Bahcall}(1997)}]{waxmanbahcall97}
{Waxman} E., {Bahcall} J., 1997, Physical Review Letters, 78, 2292

\bibitem[{{Waxman} \& {Bahcall}(1999)}]{wb99}
{Waxman} E., {Bahcall} J., 1999, PhysRevD, 59, 023002

\bibitem[{{Zhang} \& {Kumar}(2013)}]{zhangkumar13}
{Zhang} B., {Kumar} P., 2013, Physical Review Letters, 110, 121101

\bibitem[{{Zhang} \& {M{\'e}sz{\'a}ros}(2004)}]{zhangmeszaros04}
{Zhang} B., {M{\'e}sz{\'a}ros} P., 2004, International Journal of Modern
  Physics A, 19, 2385

\end{thebibliography}
\section*{Appendix A: Proton energy loss rate due to photopion interactions with a grey-body photon field}
The differential number density of a grey-body photon field is given
by 
\eqb
\nph(x) = n_0 \frac{x^2}{e^{x/\Theta}-1}, 
\eqe
where $x$ and $\Theta$ are the photon energy and the effective temperature in units of $\mel c^2$, while the normalization
$n_0$ is related to the photon compactness $\ell_{\gamma}$ as
\eqb
n_0 = \frac{15 \ell_{\gamma}}{\pi^4 \Theta^4 \sth \rb},
\label{n0}
\eqe
where $\rb$ is the size of the emission region and $\ell_{\gamma}$ does not violate the black-body limit.
The fractional energy loss rate of a proton with energy $E_{\rm p}=\gp \mpr c^2$
due to photopion interactions with the above photon field is then given by (see e.g. WB97):
\eqb
t_{\rm p\gamma}^{-1} & \equiv & -\frac{1}{E_{\rm p}}\frac{d E_{\rm p}}{d t} \large\vert_{\rm p\gamma} = \\
& = & \frac{c}{2 \gp^2}
\int_{\epsilon_{\rm th}}^{\infty}\!\!\!{\rm d}\epsilon \ \epsilon \sigma_{\rm p\gamma}(\epsilon) \xi_{\rm p\gamma}(\epsilon) \int_{\epsilon/2\gp}^{\infty}\!\!\! {\rm d}x x^{-2} \nph(x),
\label{lossrate}
\eqe
where $\epsilon$ is the photon energy (in $\mel c^2$ units) as seen in the rest frame of the proton, $\epsilon_{\rm th}$ is the threshold energy
of the interaction in $\mel c^2$ units, i.e. $\epsilon_{\rm th}\simeq 300$, $\sigma_{\rm p\gamma}$ and $\xi_{\rm p\gamma}$
are the cross section and the inelasticity of the interaction, respectively. Here, we approximate the cross section by a step function, i.e.
$\sigma_{\rm p\gamma} \approx \sigma_0 H(\epsilon-\epsilon_{\rm th})$ with $\sigma_0 = 5\times 10^{-28}$~cm$^{-2} \simeq (5/6\times10^{-3})\sth$. 
Although the inelasticity may vary between $\sim 0.1$ and 0.5 for interactions taking place 
close to and far from the threshold, respectively, for our purposes
it is sufficient to use an average value, e.g. $\bar{\xi}_{\rm p\gamma} \approx 0.2$.

\begin{figure}
 \centering
 \includegraphics[width=0.5\textwidth]{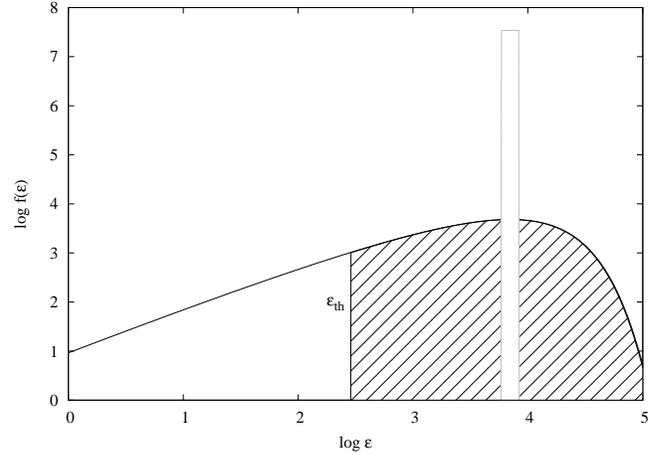}
  \caption{Function $f=-\epsilon \ln\left(1-e^{-\epsilon/\epsilon_0}\right)$ for $\epsilon_0=10^4$ (black line) and
  the $\delta$-function approximation (grey line). The hatched area denotes $I$  while the white colored area below the delta
  function corresponds to $I_{\rm appr}$.}
  \label{integral}
\end{figure}

Using these approximations and after performing the integration over $x$, eq.~(\ref{lossrate}) simplifies into 
\eqb
t_{\rm p\gamma}^{-1} = -\frac{c \Theta n_0 \sigma_0 \bar{\xi}_{\rm p\gamma} }
{2 \gp^2} \int_{\epsilon_{\rm th}}^{\infty}{\rm d}\epsilon \ \epsilon \ln\left(1-e^{-\epsilon/2\gp\Theta}\right).
\eqe
The function that appears in the above integral, i.e. 
$f(\epsilon) = -\epsilon \ln\left(1-e^{-\epsilon/2\gp\Theta}\right) $, is a steep function of $\epsilon$ that shows a sharp peak at approximately
$\ep \simeq 1.4\gp \Theta$. Hence, its integral $I$ can be approximated by $I_{\rm appr} \approx  \ep f(\ep)$, where  $f(\ep) \approx 0.7 \ep$. This
is illustrated in Fig.~\ref{integral}, where the hatched area that corresponds to the value of $I$ is approximated
by the area below the $\delta$-function centered at $\ep$. We note here that the sharpness of the peak at $\epsilon \simeq \ep$ 
is not evident because of the logarithmic scale.
If we compare $I_{\rm appr}$ with the value $I$ obtained after numerical integration of the integral, we find that their ratio 
$f_{\rm cor}=I/I_{\rm appr}$ can be modeled as
\eqb
 f_{\rm cor} \simeq 3.5 \tanh\left(\frac{\ep}{\epsilon_{\rm th}} \right),
 \eqe
for $\ep \ge \epsilon_{\rm th}$. In what follows, we will incorporate this `correction' factor into our analytical expression. 
Thus, the fractional energy loss rate for protons with Lorentz factor
$\gp > \epsilon_{\rm th}/1.4 \Theta$ is given by
\eqb
t_{\rm p\gamma}^{-1} \approx 5\times 10^{-3} f_{\rm cor}\frac{\ell_{\gamma}}{4 \pi^4 t_{\rm d} \Theta},
\eqe
where we have used eq.~(\ref{n0}) and $t_{\rm d}=\rb/c$. When normalized with respect to $t_{\rm d}$, we find
\eqb
\tau_{\rm p\gamma}^{-1} \approx 1.8 \times 10^{-5} f_{\rm cor}\frac{\ell_{\gamma}}{\Theta}, 
\eqe
which, besides the correction factor, depends only on $\ell_{\gamma}$ and $\Theta$.
\section*{Appendix B: Maximum proton energy}
The maximum energy of protons is typically calculated by balancing the acceleration and 
energy loss rates. An independent constraint comes from the so-called Hillas criterion \citep{hillas84} according to which,
the gyroradius of the most energetic particles should not exceed the size of the emission region.
Here we compare the different upper limits on the maximum proton energy.

In the simplest scenario protons are accelerated at a rate which 
is inverse proportional to their energy, namely $t^{-1}_{\rm acc}(\gamma) = \kappa e B c / \gamma \mpr c^2$, where
$\kappa$ is an efficiency factor. For simplicity, we assume $\kappa=1$.
As long as proton acceleration takes place at small Thomson optical depths, the main energy loss processes that act
competitive to the acceleration process are synchrotron radiation and photopion interactions. Note that for
larger optical depths pp-collisions consist another important energy loss mechanism for protons as well as another
neutrino production channel (e.g. \citealt{murase08}).

The synchrotron proton loss timescale is given by $t_{\rm syn} = 6 \pi \mpr c \chi^2 / \sth B^2 \gamma$, where $\chi=\mpr/\mel$. By equating
$t_{\rm acc}$ and $t_{\rm syn}$ and by using eq.~(\ref{B}) and $\rb \approx c\Gamma \dt$ we find the saturation Lorentz factor because of synchrotron losses
\eqb
\gamma_{\rm syn} \simeq 9\times10^8 \  \Gamma_2^{3/2} \dt_{-1}^{1/2} L_{\rm j, 52}^{-1/4} \epsilon_{\rm B, -1}^{-1/4}.
\label{syn}
\eqe
For the benchmark case of $\Gamma=225$, $\dt \approx 0.15$~s, $\eb=2.2\times 10^{-3}$ and 
$L_{\rm j} = 3\times 10^{52}$~erg/s, we find $\gamma_{\rm syn} \simeq 7\times 10^9 > \gamma_{\rm H}$.
 Even for $B=1.5\times 10^4$~G (see \S 4.4.1) the relation $\gamma_{\rm syn} > \gamma_{\rm H}$ holds.
 
For large values of the  photon compactness,  proton cooling due to photopion production may overcome synchrotron losses. 
The respective loss rate for protons having energies above the threshold is constant and is given by
eq.~(\ref{anal}); for the derivation, see Appendix~A. Setting $t_{\rm acc}=t_{\rm p\gamma}$ we find the saturation Lorentz factor to be
\eqb
\gamma_{\rm p\gamma} \simeq 2\times 10^9 \ \Gamma_2^2 \dt_{-1} \epsilon_{\rm B, -1}^{1/2} L_{\rm j, 52}^{1/2} L_{\gamma, 52}^{-1} \frac{E_{\rm br, obs}}{0.5 {\rm MeV}} (1+z),
\label{photopion}
\eqe
where we replaced the effective temperature $\Theta$ in eq.~(\ref{anal}) by the peak energy of the photon spectrum $E_{\rm br, obs}$. 
For the benchmark parameter values we find $\gamma_{\rm p\gamma} \approx 3\times 10^{9} (1+z)/ L_{\gamma, 52} > \gamma_{\rm H}$.  

Summarizing, for the present analysis it is safe to assume that $\gmx = \gamma_{\rm H}$.

\end{document}